\documentstyle[12pt,doublespace]{article}
\begin{document}
\begin{center}
\section*{\bf Relativity as the quantum mechanics of space-time
measurements}
 
\vspace{3mm}

Richard Lieu

Department of Physics, University of Alabama, Huntsville,
AL 35899, U.S.A.

email: lieur@cspar.uah.edu

\end{center}

\newpage

\noindent
{\bf Abstract}

Can a simple microscopic model of space and time demonstrate Special
Relativity as the macroscopic (aggregate) behavior of an ensemble ?
The question will be investigated in three parts.  First, it is shown
that the Lorentz transformation formally stems from
the First Relativity Postulate (FRP)
{\it alone} if space-time quantization
is a fundamental law of physics which must be included as part of
the Postulate.
An important corollary,
however, is that when measuring
devices which carry the basic units of lengths and time (e.g.
a clock ticking every time quantum) are `moving' uniformly, they appear
to be measuring with larger units.  Secondly, such an
apparent increase in the sizes of the quanta can be attributed to
extra fluctuations associated with motion, which are
precisely described in terms of a thermally agitated
harmonic oscillator by using a temperature parameter.  This
provides a stringent constraint
on the microscopic properties of flat space-time: it is an array
of quantized oscillators.  Thirdly, since the foregoing development
would suggest that the space-time array of an accelerated frame cannot 
be in thermal equilibrium, (i.e. it will have a distribution of temperatures),
the approach is applied to the case of acceleration by the field of {\it any}
point object, which corresponds to a temperature `spike' in the array.
It is shown that the outward transport of energy by phonon conduction
implies an inverse-square law of force at low speeds, and the
full Schwarzschild metric at high speeds.
A prediction of the new theory is that when two inertial observers
move too fast relative to each other, or when fields are too strong,
anharmonic corrections will modify effects like time dilation,
and will lead to asymmetries which implies that the FRP may not
be sustainable in this extreme limit.

\newpage 
 
In contemporary physics the FRP
is usually viewed
as a portrait of the complete symmetry between inertial observers.
Specifically all such observers experience the same laws of
physics$[1]$.  This Postulate, though simply stated, is
powerful because application of it to new
physical principles could lead to important
consequences.
The supreme example is when Maxwell's equation of electromagnetic wave 
(i.e. light)
propagation finally became recognized as a universal law of
nature$[2]$.  In order that the equation be invariant with
respect to all inertial frames, their coordinates cannot
be related by the Galilean transformation (which assumes
space and  time are absolute), as first realized by Lorentz$[3]$.
A way of arriving at the correct transformation
simply and logically, as suggested by
Einstein$[4]$, is to introduce the Second Relativity Postulate
(hereafter the SRP).  It
states that the speed of light is a universal constant, independent
of (a) inertial frames, and (b) the motion of
the emission source in each frame.  Since these two independences,
especially (b), do not arise unambiguously from the
assertion that Maxwell's equations must be formally invariant in
all frames, the need for an explicit declaration as a postulate
is inevitable. 
Thus, although
Einstein did fruitfully integrate
Maxwell's theory with the FRP, it was
necessary for him to enlist a separate postulate.
Does this mean the statement of
symmetry provided by the FRP, though very elegant, only plays the role
of compelling and guiding us to invoke further postulates
everytime we have to deal with a new physical law ?

Here I suggest another possibility, viz. that the reason for the 
problems of the FRP
is because, unlike Newton's laws, 
neither Maxwell's theory nor
Einstein's SRP operate on a sufficiently fundamental level.
The missing piece is simply that microscopically
space and time
are quantized: there exists naturally imposed units of
distances and time intervals as characterized by
the parameters $(x_o, t_o)$, the
minimum uncertainties in
one's ability to measure the coordinates $(x,t)$ of an event,
which cannot be surpassed irrespective
of an instrument's accuracy.
When incorporated with the FRP,
the Postulate now reads: {\it the
laws of physics as observed through a system of quantized
space-time,} the latter a `grid' which inevitably
controls the outcomes of measurements,
{\it is the same for all inertial observers}.
Thus, if a frame transformation
arbitrarily changes $x_o$ and 
$t_o$, the reference `grid'
will in general be distorted; but if at
least the ratio of the two quantities
remains constant, the `grid' will be enlarged or reduced
proportionately, and so long as no physics exists which depends
on absolute scales, such a transformation will still preserve
the symmetry between inertial frames.
It is worth investigating whether this generalized FRP {\it alone} will lead
to a unique set of
equations which connect the coordinates of different frames,
Before doing so, however,
I designate the ratio $x_o/t_o$ a universal constant $v_o$ which
has the dimension of a speed.

Let us first examine the Galilean transformation $G$. The
event data of an inertial
observer $\Sigma$ are $(x,t)$ with
accuracies $(x_o,t_o)$.  According to $G$, such data would appear
differently to
another observer $\Sigma'$ who moves relative to $\Sigma$ with
velocity ${\bf v} = (v,0,0)$.  More precisely this observer 
notices that if he were to repeat
the measurement following every step adopted by $\Sigma$, his
results would be $(x',t')$ with accuracies $(x'_o,t'_o)$,
where $x' = x - vt$,
$t' = t$; and $x'_o = x_o \sqrt{1 + v^2/v_o^2}$, $t'_o = t_o$.
Based on the above perception, $\Sigma'$ constructs a physical
model of his world view.  Would this be the same model as the one
derived from experiments performed directly by himself ?  Such a
question is indeed valid and pertinent
if we interpret the forementioned accuracies as
{\it limiting} ones which form basic units of measurements.  The
answer is `no' because of two
problems: (a) the measurement `grid'
is distorted because the ratio
$x_o/t_o$ changes
from $v_o$ to $v'_o = x'_o/t'_o = v_o \sqrt{1 + v^2/v_o^2} > v_o$;
(b) a consequence of space-time quantization is that measured
distances transform if and only if the unit
of distance does (the same applies to time intervals); in this
regard $G$, which clearly preserves distances but changes
$x_o$ to $x'_o > x_o$, is not even a self-consistent transformation.

I proceed to seek a frame transformation which conserves the
ratio $x_o/t_o$.
We start with $\Sigma$ and
$\Sigma'$ having their spatial (cartesian) coordinate axes parallel
to each other, and common space-time origin.  The 
macroscopic homogeneity and
isotropy of space-time, 
a corollary
of the FRP,
immediately implies a linear relationship between the two sets
of 4-D coordinates with $y$ and $z$ separated from the rest, i.e.
\begin{eqnarray}
t' = r(v) [t - s (v) x],~~~x' = q(v) [x - p(v) t],~~~y'=y,~~~z'=z \nonumber
\end{eqnarray}
where the functions $q$, $r$, and $s$ can depend only on $v$.
By taking into account the fact that
an object stationary relative to $\Sigma'$ moves
relative to $\Sigma$ with
speed $v$ along $+x$,  
one easily deduces $p(v) = v$, so that
\begin{equation}
t' = r(v) [t - s (v) x],~~~x' = q(v) (x - vt),~~~y'=y,~~~z'=z
\end{equation}
Next,
since the combined operations ${\bf r} \rightarrow {\bf r'}$,
$t \rightarrow t'$ and $v \rightarrow -v$ leave the situation
unchanged, and since $q$ and $r$ are positive definite scale factors
which cannot be sensitive to the sign of $v$ (else there would
be a preference between $+x$ and $-x$), the inverse transformation
reads:
\begin{equation}
t = r(v) [t' - s (-v) x'],~~~x = q(v) (x' + vt'),~~~y=y',~~~z=z'
\end{equation}
If in the x-transformation part of equ (1) one substitutes
the expressions for $x$ and $t$ from equ (2), and the result
were to hold for all times, the coefficient of $t'$ must vanish.
This means:
\begin{equation}
q(v) = r(v)
\end{equation}

The development thus far has been based on general considerations,
the functions $q(v)$
and $s(v)$ remain arbitrary.  Einstein$[4]$ determined them
by applying the SRP: the speed of a light signal 
as measured by the two observers must be
equal, i.e. $\Delta x/\Delta t = \Delta x'/\Delta t'= c$, where,
from Equs (1) and (3), $\Delta t' = q(\Delta t - s \Delta x)$ and
$\Delta x' = q(\Delta x - v \Delta t)$.  This leads to $s = v/c^2$,
and the criterion of a consistently invertible transformation
restricts $q$ to:
\begin{equation}
q(v) = \frac{1}{\sqrt{1-\frac{v^2}{c^2}}},
\end{equation}

What happens if, instead of using the SRP, we apply the constancy
of $x_o/t_o = v_o$ ?  Here the two approaches differ, because unlike
$\Delta x$, the uncertainties of measurements, $\delta x$,
do not transform linearly.  If $\Sigma$ measures
$(x,t)$ independently to accuracies $\delta x$ and $\delta t$
then, defining the primed quantities as before, we have:
\begin{equation}
(\delta x')^2 = q^2 [ (\delta x)^2 + v^2 (\delta t)^2];~~~
(\delta t')^2 = q^2 [ (\delta t)^2 + s^2 (\delta x)^2] 
\end{equation} 
For the limiting (quantum) uncertainties the above formulae 
remain valid with
the substitutions:
\begin{equation}
\delta x = x_o,~~\delta x' = x'_o,~~\delta t =t_o,~~\delta t'=t'_o
\end{equation}
The
requirement $\delta x'/\delta t' = \delta x/\delta t = v_o$ implies
that $s = \pm v/v^2_o$, i.e. {\it two classes of transformation
are permitted}.  Now the criterion of invertibility restricts $q$,
in the case of the first solution, to the form:
\begin{equation}
q(v) = \frac{1}{\sqrt{1-\frac{v^2}{v_o^2}}}~~for~~s = \frac{v}{v_o^2}
\end{equation}
or, as second solution, to:
\begin{equation}
q(v) = \frac{1}{\sqrt{1+\frac{v^2}{v_o^2}}}~~for~~s = - \frac{v}{v_o^2}
\end{equation}

It is easily verified that the second solution re-scales
distances and time intervals, but does not
transform the quantities
which determine units of measurement, i.e. $x'_o$ remains
equal to $x_o$ etc.
Thus, by applying the reason (b) given above (which explained why $G$
is not an acceptable transformation), we can likewise exclude this second
solution.  The first solution, of course, has
the mathematical form
of the Lorentz transformation $L$ of Special Relativity, even though we
evidently undertook a different path to obtain it.  Here the re-scaling 
of coordinates is entirely consistent with changes in the
sizes of the two quanta, as I shall demonstrate.
Moreover, since it is clear that
there is only one known speed
sufficiently universal to participate in the
present consideration, viz. the speed of light in vacuum $c$,
I shall without further arguments set:
\begin{equation}
v_o = c
\end{equation}
and our
results are identical to $L$.

Is such an alternative approach to
Special Relativity
a mere pedagogy ?
A crucial difference from
Einstein's theory is that a further elucidation of the transformation
scale factor $q$, the well known Lorentz factor, is now available.
When the
positioning and timing measurements of a `moving' observer are
recorded from a `stationary' reference frame, the former
appears to have adopted larger basic units.  The phenomenon can
completely be explained by quantizing space-time
as harmonic oscillators (cf. black body radiation).
To prove this important
point, we remind ourselves of the intrinsic uncertainties
$x_o$ and $t_o$ which an experiment performed aboard any inertial
platform is
subject to.
Now an observer notices that his partner in relative motion
suffers from the higher uncertainties $x'_o$ and $t'_o$.
I shall only examine $x'_o$, {\it since the treatment
of $t'_o$ is analogous}.  The first part of equ (5) may be re-written,
with the help of equs (6), (7), and (9), as:
\begin{equation}
x_o^{'2} = x_o^2 \left(1+\frac{\frac{2 v^2}{c^2}}{1 - \frac{v^2}{c^2}} \right)
\end{equation}
It is reasonable to correspond the minimum value, $x_o$, with
the width of the position distribution of a 1-D harmonic oscillator
at ground state (which is a gaussian).
Denoting the zero-point energy and oscillator
constant by $\epsilon/2$ and $\kappa$, respectively,
we have:
\begin{equation}
\epsilon = 2 \kappa x_o^2
\end{equation}

I consider such a standpoint because of its remarkable interpretation
of equ (10).  
Suppose motion enlarges $x_o$ because at finite $v$ an oscillator
can populate the excited states, the degree of which is expressed by a
temperature parameter $T$, just like a system in thermal equilibrium
($T$ increases with $v$ and $T=0$ when $v=0$).
We may then write
$x_o^{'2} = x_o^2 (1+<x_1^2>/x_o^2)$,
where $<x_1^2>$ is the mean-square amplitude due to the occupation of
all energy levels higher than the zero-point, and is
related to the mean energy of these upper levels,
\begin{equation}
\bar{E} = 
\frac{\epsilon e^{-\frac{\epsilon}{kT}}}{1 - e^{-\frac{\epsilon}{kT}}},
\end{equation}
by $\kappa <x_1^2> = \bar{E}$.  Therefore:
\begin{equation}
x_o^{'2} = x_o^2 \left(1 + \frac{\epsilon}{\kappa x_o^2}
\frac{e^{-\frac{\epsilon}{kT}}}{1-e^{-\frac{\epsilon}{kT}}} \right)=
x_o^2 \left(1 + \frac{2 e^{-\frac{\epsilon}{kT}}}{1-e^{-\frac{\epsilon}{kT}}}
\right) 
\end{equation}
Equs (10) and (13), when taken together, suggest strongly the
following association:
\begin{equation}
\frac{v^2}{c^2} = e^{-\frac{\epsilon}{kT}}~~
\left(= \frac{\bar{E}}{\epsilon + \bar{E}} \right).
\end{equation} 
The two equations are in perfect agreement.
Thus the behavior of $x_o$ during uniform motion has an exact
parallel with the
fluctuation enhancement of a thermally agitated 
quantum oscillator.  In this way,
we bolster {\it a posteriori} the earlier premise of
space-time quantization.  Moreover, there is now a clear rationale for
postulatin the assignment of a temperature to the
space-time array of a moving observer, the ground-breaking potential
of such an undertaking will become apparent when we address
non-inertial behavior.  We note also that the thermodynamics$[5]$ and
quantum mechanics of space-time has been a subject of much
investigation (see, e.g., the recent review of Ashtekar[6]), even
though this is the first direct attempt in explaining Relativity as
macroscopic (aggregate) behavior of a simple quantum ensemble.

Before doing so, I propose the following
microscopic model
of space-time.  It is a 4-D array of `nodes' (or `measurement tickmarks'),
all adjacent pairs of which are connected by identical harmonic
oscillators with natural length equal to the minimum fluctuation 
$x_o = c t_o$, where $x_o$ is
given by equ (11) (this reflects my earlier indication that the grid
structure is controlled by the intrinsic quantum uncertainties).
However, even at $v=0$, zero-point fluctuations are inevitable, and
enlarges the separation to $x_m = \sqrt{2} x_o$, which forms the unit
of distance for the rest observer $\Sigma$.  When there is relative
motion ($v > 0$ and hence $T > 0$) $x_m$ increases to:
\begin{equation}
x_m^{'2} = x_m^2 (1+\frac{<x_1^2>}{x_m^2})
\end{equation}
where, as before, 
$\kappa <x_1^2> = \bar{E}$ and $\bar{E}$ is given by equ (12).
Thus we have:
\begin{equation}
x'_m = x_m \left(1 + \frac{\epsilon}{2 \kappa x_o^2}
\frac{e^{-\frac{\epsilon}{kT}}}
{1-e^{-\frac{\epsilon}{kT}}}\right)^{\frac{1}{2}} =
\frac{x_m}{\sqrt{1-\frac{v^2}{c^2}}}
\end{equation}
This proves that our first solution of the FRP, equ (7), is
fully self-consistent.  Specifically the re-scaling of coordinates
by the Lorentz factor $q$ (equs (7) and (9))
is due solely to the change of natural units with motion,
thereby satisfying the basic requirement that a robust space-time
quantization scheme should be compatible with the FRP.

The idea of inertial space-time arrays being equilibrium 
configurations (albeit having different temperatures) is further
strengthened by its application to the problem of acceleration,
because logical deduction would suggest this should then
correspond to a situation where the array is out of equilibrium,
and is characterized by a distribution of temperatures.
Consider the simple case of a point (delta-function) enhancement
in the temperature at the origin of a rest array, which leads to
the {\it isotropic} conduction of energy in all four directions
$(ct, {\bf r})$.

Again, I will only solve for one spatial
coordinate $x$, as the other three will follow in a likewise manner
(with $ct$ replacing $x$ for the case of time).  The x-axis is
obviously a radial direction, and we let the energy propagate
outwards from some minimum radius $x_{min}$.  The transport equation is:
\begin{equation}
- \sigma_{th} \frac{dT}{dx} = n \bar{E} \bar{v}.
\end{equation}
Here $x$ is the distance from $x = x_{min}$ to any `downstream' point,
measured, of course, using the oscillator lengths
at all intermediate points, which are no longer
uniformly distributed, as basic units (we shall find that $x$ is
after all an Euclidean distance).  Moreover,
$\sigma_{th} = n \bar{v} \lambda d\bar{E}/dT$ is the
thermal conductivity of phonons: $n$ is the linear phonon density
(i.e. always one oscillator per unit length)
$\bar{v}$ is the `speed of sound' in the array\footnote{The conduction
of energy takes place throughout the entire space-time array.
Thus, like the oscillations,
there is the need to introduce a new `time' axis when defining
the propagation speed of these phonons -
signature of a fifth dimension.}, $\bar{E}(T)$ is the mean energy
of a phonon above the ambient (zero-point) energy,
and $\lambda$ is the phonon mean free path
measured in the same way as $x$ is.  Since phonons do not interact,
this is simply
the size of the available array,
i.e. $\lambda = x$.  Thus equ (17), together with the meaning of
the various symbols involved\footnote{Equ (17), which
is a standard heat transport equation, may be tested by applying it
to a homogeneous ideal classical gas.  In this case $n$ and $\bar{v}$ are
respectively the number density and mean speed of the gas
particles, and $\bar{E} = \frac{3}{2} kT$.  Also, unlike
phonons, the mean free path is in general much smaller than
the total dimension occupied by the gas.  Solution of equ (17)
then readily leads to the well-known result that the
e-folding length for the temperature profile is $\lambda$.},
imply that
\begin{equation}
-x \frac{d\bar{E}}{dx} = \bar{E},~~or~~ \bar{E} = \frac{1}{\alpha x}
\end{equation}
where $\alpha$ is a constant of integration.  Combining equs (14) and (18),
one obtains
\begin{equation}
\frac{v^2}{c^2} = \frac{1}{\alpha \epsilon x +1}
\end{equation}
The oscillators indeed have a `profile' of lengths (reflecting a
temperature drop with distance) which is reproduced by assigning to
every point $x$ a local Lorentz frame moving at speed $v$.  Evidently
$v$, and hence the enlargement of the length quantum, decreases with 
$x$.  This is because the temperature returns gradually to the ambient
value of $T=0$ as energy is transported downstream.

Now since $x$ is a radial distance we may write
$x = r - r_o$ where $r_o = x_{min}$.  In the limit
of $r \gg r_o$ ($v \ll c$) we have $v^2 \propto 1/r$.
A careful reader will
realize that the situation is the same as either (a) a point mass
at the origin causing space-time curvature which attracts all
other masses inwards (equ (19) gives the free-fall speed at every
position), or (b) the observer is in an accelerated frame, responding to
non-gravitational forces which act along the $+x$ direction, due again
to the fields of a point object.  The Principle of Equivalence excludes
the absolute certainty of distinguishing between the two possibilities.
This means, for the first time, one can {\it derive} from more
fundamental principles that {\it the far-field potential of any field
emanating from a point source is $\propto 1/r$, or the force obeys
inverse-square law}.

Further, in the case of (a), agreement with Newtonian
gravity is obtained by setting $\alpha \epsilon = c^2/2GM$, which removes
the arbitrariness of the solution.  
At high speeds the role of $r_g$ must be taken into
account ($v \rightarrow c$ as $r \rightarrow r_o$).
In the present case of spherical symmetry there is only one
free parameter in the problem, i.e. $r_g$ must depend
on $\alpha$.  By setting $r_o = 1/\alpha \epsilon = 2GM/c^2$, equ (19)
reduces to $v^2/c^2 = r_o/r$, or $q = (1 - \frac{r_o}{r})^{-\frac{1}{2}}$.
If we bear in mind that the outward energy conduction is isotropic
in the {\it space-time} array, i.e. the temperature of
the oscillators in the $ct$ direction 
is distributed identically to the $r$ direction, it will become obvious 
that the space-time units are constantly decreasing as one moves
away from the origin.
When this set of variable units is used to measure
distances and time intervals, as we did, we have in fact adopted
an Euclidean geometry whereby the path of all freely moving objects
is a straight line.  The above expression for the Lorentz factor
$q$ contains all the information
one needs to construct the full Schwarzschild metric$[7]$
of General
Relativity.  For example, as a falling object approaches
the gravitational radius (or event horizon) $r = r_o$
time dilation becomes infinite.  The 
forementioned notion of a minimum conduction radius
is now clear: phonon energy transport to other parts of
the array takes place only beyond the event horizon.  Following the
earlier arguments, we realize that {\it this relativistic correction 
applies to point interaction involving any type of fields}, which
is also a new conclusion.

Lastly I propose a possible test of the theory.  The domain
within which the quanta of space-time oscillations manifest
themselves collectively as Special Relativistic effects is
the `harmonic limit'.  It is well known that the forces which
maintain stability of a system
may always be approximated by a harmonic oscillator potential in the
case of small perturbations about an equilibrium point.
If an oscillator's temperature is too high (meaning here that
$v$ is too close to $c$) anharmonic terms will no longer be
negligible, and will correct the Lorentz transformation.
Here I inquire the form by which time dilation could
be modified.  

According to equ (15), the increase of a
measurement unit with $v$ is due to 
the term
$<x_1^2>$, which in the
classical (high T) limit may be written as:
\begin{equation}
<x_1^2> = \frac{\int x^2 e^{- \beta V(x)} dx}{\int e^{- \beta V(x)} dx}
\end{equation}
where $\beta = 1/kT$ and 
\begin{equation}
V(x) = \frac{1}{2} \kappa x^2 + \xi x^3 + \eta x^4
\end{equation}
Hitherto all but the first term of $V(x)$ have been ignored.  However,
in the Taylor series of equ (21) the coefficients
are successive derivatives of $V(x)$.  For a regular array of nodes, it
is therefore reasonable to assume, like solid state theory,
that $\kappa/\xi$ is comparable to
$\xi/\eta$ etc.  In this way one can include all contributions
which belong to the next order of small quantities, but not beyond.
The result is:
\begin{equation}
\frac{1}{2} \kappa <x_1^2> = \frac{1}{2} kT + \frac{1}{2} \alpha (kT)^2
\end{equation}
where $\alpha = 45 \xi^2/\kappa^3 - 12 \eta/\kappa^2$.
Substituting equ (22) into equ (15), and applying equ (11), one
obtains the ratio:
\begin{equation}
\frac{x'_m}{x_m} = \frac{1}{\sqrt{1 - \frac{v^2}{c^2}}}
\left(1 + \frac{\alpha \epsilon}{1 - \frac{v^2}{c^2}} \right)^{\frac{1}{2}}
\end{equation}
which predicts that at very high speeds time dilation
deviates from a simple dependence on the Lorentz factor.

To date the best direct experiments on time dilation
at large $v$ remains those which measure the integral energy spectrum 
of vertical muons at sea level, where the steepening of the
power-law index by unity as compared with that of the parent
pions and protons is in agreement with
Special Relativity$[8]$ up to an energy of 10 TeV$[9]$, or muon Lorentz
factor $\sim$ 10$^5$.  This constrains the 
magnitude of the coefficient of the correction term to
$\alpha \epsilon <$ 10$^{-10}$.  Further, equ (23) will have a
`feedback'
effect on the Lorentz transformation and the FRP.  A thorough
analysis will be performed in a later work, except to say here that
the statement of complete symmetry provided by the
FRP might not be sustainable at such high values
of $v$ because when the $\alpha \epsilon$ term is not
negligible equ (23) confounds the linearly invertible property
of $L$.
This, an issue which has been raised before$[10]$, can
be treated quantitatively using the present formalism.  It is
also a limit which can be investigated 
by observing extreme energy
cosmic ray neutrino events at Lorentz factors $>$ 10$^{20}$, where
the $\nu + p$ interaction at $E_{\nu} =$ 10$^{20}$ eV is equivalent to
the process of $\nu + p$ at $E_p = 10^{29}$ eV.  In the latter case
the de Broglie wavelength of protons is smaller than the Planck
length by more than two orders of magnitude, yet the same effect is
not as obvious in the former case.  This
highlights
the possibility of an asymmetry in the frames of reference represented
by the two cases.

In conclusion, it is argued that Relativity is the macroscopic
manifestation of space-time as a 4-D array (ensemble)
of quantized harmonic oscillators.
The Lorentz transformation, which connects the space-time
arrays of different inertial frames,
simply `maps' the nodes of two equilibrium oscillator ensembles
which have a relative temperature difference between them.
Accelerated frames, however, are no longer associated with equilibrium
states, but rather such ensembles have a temperature gradient within them.
As an example, for a `point-
enhancement'
the result is in complete agreement with General Relativity, with the
inverse-square law of force as a limiting case.  The theory can cope
with many new situations, 
including a prediction of how time dilation might be modified at
very high speeds.
Other possibilites are changes in the structure of
space-time near the event horizon (where oscillations are not
harmonic) which may affect the gravitational bending of
light, and diffraction of very energetic photons (with wavelengths
comparable to $x_o$) by the space-time lattice.

I gratefully acknowledge
L. Hillman, Y. Takahashi, M. Bonamente and R.B. Hicks
for very helpful discussions and critical reading of the
manuscript.
I am also indebted to W.I. Axford,
R.D. Blandford, and J. Malenfant (the editor) for their
encouragement.

\newpage

\noindent
{\bf References}

\noindent
$[1]$ Poincare, H., 1905, `The Principles of Mathematics',
{\it The Monist}, {\bf 15}, 1 -- 24. \\
\noindent
$[2]$ Hertz, H., 1893, in {\it Electric Waves} (translation
by D.E. Jones), London Macmillan;\\
\indent
reprinted by Dover Pubs., Inc., 1962. \\
\noindent
$[3]$ Lorentz, H.A., 1904, `Electromagnetic phenomena in a
system moving with any \\ 
\indent velocity less
than that of light', {\it Proc. Amsterdam Acad Sci},
{\bf 6}, 809 -- 831. \\
\noindent
$[4]$ Einstein, A., 1905, {\it Annalen der Physik}, {\bf 18}, 891.\\
\noindent
$[5]$ Bardeen, J.W., Carter, B., Hawking, S.W., 1973, {\it Commun
Math Phys}, {\bf 31}, 161 -- 170. \\
\noindent
$[6]$ Ashtekar, A., 2000, {\it Annalen Phys}, {\bf 9}, 178 -- 198. \\
\noindent
$[7]$ Schwarzschild, K., 1916, \"{U}ber das Gravitationsfeld
eines Massenpunktes \\
\indent
nach der Einsteinschen Theorie, {\it Sitzber Preuss. Acad. Wiss. Berlin}\\
\indent
189 -- 196. \\
\noindent
$[8]$ Hayakawa, S., Nishimura, J., Yamamoto, Y., 1964, {\it Supplement of the\\
\indent Progress of Theoretical Physics}, {\bf 32}, 104 -- 153. \\
\noindent
$[9]$ Mizutani, K., Misaki, A., Shirai, T., Watanabe, Z., Akashi, M.,\\
\indent Takahashi, Y., 1978, {\it Il Nuovo Cimento}, {\bf 48}, 429 -- 445.\\
\noindent
$[10]$ Coleman, S., Glashow, S., 1997, {\it Phys Lett}, {\bf 47}, 1788. \\

\end{document}